\documentclass[%
 aip,
 amsmath,amssymb,
 reprint,%
]{revtex4-1}
\usepackage{graphicx}
\usepackage{dcolumn}
\usepackage{bm}
\usepackage[utf8]{inputenc}
\usepackage[T1]{fontenc}
\usepackage{mathptmx}
\usepackage{etoolbox}
\usepackage{url}

\usepackage{breakurl}
\usepackage[breaklinks]{hyperref}
\usepackage{xcolor}

\makeatletter
\def\@email#1#2{%
 \endgroup
 \patchcmd{\titleblock@produce}
  {\frontmatter@RRAPformat}
  {\frontmatter@RRAPformat{\produce@RRAP{*#1\href{mailto:#2}{#2}}}\frontmatter@RRAPformat}
  {}{}
}%
\makeatother
\begin{document}

\preprint{AIP/123-QED}

\title[Noise-induced excitability: bloom, bust and extirpation in autotoxic population dynamics]{Noise-induced excitability: bloom, bust and extirpation in autotoxic population dynamics}
\author{P. Moreno-Spiegelberg}
\email{pablomorespi@gmail.com}
\affiliation{Department of Agricultural Sciences, University of Naples Federico II, 80055 Portici, Naples, Italy.}
\affiliation{IFISC (CSIC-UIB), Instituto de F\'{\i}sica Interdisciplinar y Sistemas Complejos, E-07122 Palma de Mallorca, Spain.}

\author{J. Aguilar}
\thanks{P.M.S. and J.A. contributed equally to this work.}
\affiliation{Dipartimento di Fisica e Astronomia Galileo Galilei, Universit\`{a} degli Studi di Padova, via Marzolo 8, 35131 Padova, Italy}

\date{\today}

\begin{abstract}
Species populations often modify their environment as they grow. When environmental feedback operates more slowly than population growth, the system can undergo boom-bust dynamics, where the population overshoots its carrying capacity and subsequently collapses. In extreme cases, this collapse leads to total extinction. While deterministic models typically fail to capture these finite-time extinction events, we propose a stochastic framework, derived from an individual-based model, to describe boom-bust-extirpation dynamics. We identify a noise-driven, threshold-like behavior where, depending on initial conditions, the population either undergoes a ``boom'' or is extirpated before the expansion occurs. Furthermore, we characterize a transition between an excitable regime, where most trajectories are captured by the absorbing state immediately after the first bust, and a persistent regime, where most populations reach a metastable state. We show that this transition is governed by the noise strength and the ratio of environmental-to-population timescales. This framework provides a theoretical basis for understanding irreversible transitions in invasive species, plant succession, microbial dynamics, and the elimination of cancerous tumors.
\end{abstract}

\maketitle

\begin{quotation}
Boom–bust dynamics, characterized by rapid population growth followed by an abrupt collapse, are a hallmark of nonlinear systems governed by time-lagged feedbacks. While deterministic models effectively capture these oscillations, they fail to account for the irreversible extirpation—or extinction—observed in many empirical scenarios, as deterministic trajectories only reach absorbing states asymptotically. In this work, we propose a stochastic framework, derived from microscopic first principles, to describe ``boom-bust-extirpation'' dynamics. We identify a novel mechanism termed ``noise-induced excitability'', which resembles the behavior of deterministic excitable systems but where both the excitation threshold and the final capture by the absorbing state are driven by demographic noise. By combining numerical simulations with analytical approximations, we characterize the transitions between different extinction pathways, providing a theoretical basis for predicting sudden collapses in populations ranging from invasive species to microbial communities.
\end{quotation}

\section{Introduction}

Species introduced into new ecosystems often exhibit boom–bust dynamics \cite{Strayer2017}, characterized by rapid population growth (boom) followed by an abrupt collapse to low abundance (bust). These declines are typically driven by negative feedback mechanisms, such as resource depletion \cite{aguilar2025unraveling}, toxin accumulation \cite{Piccardi2019}, or increased pressure from pathogens \cite{Stricker2016} and predators \cite{Keane2002,Carlsson2011}. While extensively studied in ecology, these dynamics are of broader interest in physics and nonlinear science because they embody universal features of systems governed by feedbacks, instabilities, and time-lagged responses.

After a bust, the factors that reduced population densities may gradually weaken—for example, resources are replenished, toxins break down, or pathogen levels fall—creating conditions that allow populations to recover. This can lead to consecutive boom-bust events. In other cases, however, the cycle occurs only once, shortly after the introduction of a species. The outcome may be a stable population at finite density or, alternatively, a complete collapse to extinction (extirpation) \cite{Simberloff2004,Lester2016}. The latter scenario, which we call \emph{boom-bust-extirpation} dynamics, is particularly striking because extinction represents an absorbing state: once there are no individuals in the system, recovery is impossible without external intervention. Understanding the mechanisms driving this irreversible transition is the central focus of this work.

Deterministic population models can capture growth and collapse, but they fundamentally fail to account for extinction at finite times. In these descriptions, the absorbing state is reached only asymptotically in the limit of infinity times. Furthermore, in deterministic models, even a vanishingly small population will recover if growth conditions are favorable, contradicting the irreversibility of extinction processes where zero population density is reached. A common ecological resolution to this limitation is to assume a strong Allee effect. The Allee effect describes a positive correlation between population size and mean individual fitness at low abundances. Mathematically, this translates into a nonlinear growth rate characterized by positive feedback within small populations. When this Allee effect is strong, the system exhibits a critical population density threshold characterized by a net zero growth rate; consequently, populations falling below this critical value inevitably decline toward (though deterministically never reaching) zero, even under optimal environmental conditions \cite{MorenoSpiegelberg2025}. While effective in certain contexts, this mechanism relies entirely on explicit positive feedback and fails to address the fundamental role of demographic stochasticity at low densities, where random fluctuations, rather than mean-field trends, dominate the system dynamics \cite{Ovaskainen2010, Aguilar_endemic}. While effective in certain contexts, this mechanism relies entirely on explicit positive feedback and fails to address the fundamental role of demographic stochasticity at low densities, where random fluctuations, rather than mean-field trends, dominate the system dynamics \cite{Ovaskainen2010, Aguilar_endemic}.

The interaction between stochastic fluctuations and nonlinear dynamics often induces novel behaviors that are absent in purely deterministic counterparts. These noise-induced phenomena have been extensively studied in the literature, spanning effects such as synchronization \cite{Shajan2023,Majhi2016}, state switching \cite{Biancalani2014}, stochastic resonance \cite{McDonnell2009}, or noise-induced chaos \cite{BassolsCornudella2023}. As we will show in this article, demographic fluctuations, naturally present in real systems but often neglected in deterministic models, can trigger the extinction of the population following a boom-bust sequence, even in the absence of an Allee effect. This result not only presents a new framework to study such events in ecology through the lens of stochastic systems, but also introduces a noise-induced excitable behavior, relevant to other fields of nonlinear science.

In this study, we propose a framework based on stochastic differential equations (SDEs) derived from first-principles individual-level behaviour \cite{VanKam}. As a case study, we examine the evolution of autotoxic populations, in which individuals produce a substance that is harmful to conspecifics~\cite{Singh01111999,Carten2016} 

Through a combination of stochastic simulations and approximate analytical treatment, we characterize the different dynamical regimes of the autotoxin system in terms of absorbing probabilities~\cite{Redner2002,Artime2018}. We show that slow toxin dynamics generate boom–bust cycles: rapid growth leads to toxin buildup and a subsequent collapse. At low densities, stochastic fluctuations can push the system into the absorbing extinction state. Because the noise amplitude scales inversely with carrying capacity (system size), large populations are likely to survive the low-density phase and recover, whereas small populations face a substantial risk of extirpation.

For the proper parameter values, we observe threshold-like behaviour and extinction after the first boom-bust cycle. We have termed this regime as \emph{noise-induced excitability}, as it resembles properties of deterministic excitability \cite{Izhikevich2006}, i.e., threshold-like response and relaxation dynamics that can be repeatedly triggered. Yet, it differs fundamentally in that both the excitation threshold and the final absorption are noise-driven. While we focus on population dynamics, this concept of noise-induced excitability could be broadly applied in neuroscience \cite{Izhikevich2006}, chemical kinetics, eradication of epidemics after an outbreak \cite{vanHerwaarden1997}, viral clearance after an acute infection, total elimination of cancerous cells in cancer treatment \cite{Lopez2019,Ramirezvaila2023}, and spatiotemporal excitability \cite{Vincenot2017}.

The paper is structured as follows: In Sec.~\ref{Sec:1}, the model and its derivation are described, and the deterministic counterpart is described. In Sec.~\ref{Sec:2}, we categorize and study the probability of the different trajectories of the system. In Sec.~\ref{Sec:3}, the results are discussed. Finally, in Sec.~\ref{Sec:5}, we introduce some concluding remarks and perspectives.

\section{The model}
\label{Sec:1}

\subsection{Microscopic derivation of mesoscopic description}\label{sec:Microscopic_derivation}
In this work, we focus on species that modify their environment by producing toxins, which, paradoxically, limit their own population expansion. This ecological process is known as ``autotoxicity'', and it occurs across a wide range of taxa and spatial scales~\cite{Singh01111999,Piccardi2019}. Ultimately, our model will describe fluctuations at a mesoscopic level of description, i.e., characterizing the evolution of species densities rather than single individuals. However, we derive such a mesoscopic description from microscopic (individual-level) first principles. In this way, we can link parameters at the mesoscopic level with those at the microscopic one, which has a more transparent biological and physical meaning, such as, for example, per-capita death rates. To do so, we follow Van Kampen's system expansion~\cite {VanKam}, which is a standard procedure within out-of-equilibrium stochastic analysis. The aim of Van Kampen's system expansion is to approximate a microscopic model — formulated as a master equation for discrete random variables — by a coarser, mesoscopic description in terms of an SDE. This is achieved by truncating a series expansion of the master equation, which yields a Fokker-Planck equation. The SDE description is then directly obtained from the latter, since the relation between Fokker-Planck equations and Langevin equations is well-known~\cite{risken_fokker-planck_1996,Gardiner1990}. In particular, our starting point is a simple agent-based model without spatial interaction, describing an autotoxicity process. Defining $N_P$ and $N_T$, the number of individuals and toxin molecules, respectively, we characterize the probability of transitions by the transition rates:
\begin{equation}
    W(\vec{N}_{t+\Delta t}|\vec{N}_{t})=\lim_{\Delta t\to 0} \frac{1}{\Delta t} \text{Prob.}\left(\vec{N}_{t+\Delta t}|\vec{N}_t \right),
\end{equation}
where $\vec{N}= (N_P,N_T)$. Specifically, the only non-zero transition rates encode asexual reproduction of individuals with rate $\beta_P$, toxin production with rate $\beta_T$, natural plant death with rate $\mu_P$, and natural toxin decomposition with rate $\mu_T$, which read
\begin{align}
    W(N_P+1,N_T|N_P,N_T) & = \beta_P \, N_P, \nonumber \\
    W(N_P-1,N_T|N_P,N_T) & = \frac{\mu_P}{V} \, N_P \,N_T, \nonumber \\
    W(M_P,N_T+1|N_P,N_T) & = \beta_T\, N_P, \nonumber \\
    W(N_P,N_T-1|N_P,N_T) & = \mu_T  \,N_T. 
\end{align}
The above rates describe constant per-capita birth rates for both individuals and toxins, representing asexual reproduction of individuals and constant toxin production, respectively. While the per-capita decay of toxins is equally constant, the death of individuals is mediated by the presence of toxins. We model the interaction between toxin molecules and individuals through a classical mass-action law: individuals in contact with toxins die at a rate $\mu_P$, while the encounter probability depends on the density of toxins $N_T/V$. The volume $V$ represents the space occupied by the population. These sorts of mass-action laws are the minimal description of encounter events in well-mixed models across chemistry, epidemics~\cite{Keeling2008}, and ecology~\cite{palamara2021}. 

The evolution of the probabilities $\mathbb{P}$ of finding the ecological system in a particular state is governed by the master equation~\cite{VanKam}:
\begin{align}\label{eq:m_eq}
    \partial_{t} \mathbb{P}(N_P,N_T) =& - \mathbb{P}(N_P,N_T)\left[\beta_P \, N_P+\frac{\mu_P}{V} \, N_P \,N_T+\beta_T\, N_P+\mu_T  \,N_T\right] \nonumber \\
    &+\mathbb{P}(N_P-1,N_T)\,\beta_P \, (N_P-1) \nonumber \\
    &+\mathbb{P}(N_P+1,N_T)\,\frac{\mu_P}{V} \, (N_P+1) \,N_T \nonumber \\
    &+\mathbb{P}(N_P,N_T-1) \, \beta_T N_P \nonumber \\
    &+\mathbb{P}(N_P,N_T+1)\, \mu_T (N_T+1)
\end{align}

By defining the densities $x=N_P/V$ and $y=N_T/V$ and expanding Eq.~\eqref{eq:m_eq} in a series of $1/V$, we find the Fokker-Planck equation for the probability density function for $x$ and $y$,
\begin{align}
    \partial_t p(x,y) =\, & \partial_x [(\mu_P\,x \,y -\beta_P\,x)\,p(x,y)] \nonumber \\
    &\partial_y [(\mu_T\,y  -\beta_T\,x)\,p(x,y)] \nonumber \\
    &+\frac{1}{2V}\partial^2_x [(\mu_P\,x \, y +\beta_P\,x)\,p(x,y)]  \nonumber \\
    &+\frac{1}{2V}\partial^2_y [(\mu_T\,y  +\beta_T\,\,x)\,p(x,y)].
    \label{Eq:FP_full}
\end{align}
Eq.~\eqref{Eq:FP_full} is compatible with the stochastic evolution of densities in terms of the Itô SDEs~\cite{VanKampen1981}
\begin{align}\label{eq:general_SDE}
    \dot{x} & = \beta_P x - \mu_P \, x \, y + \sqrt{\frac{1}{V}} \sqrt{ \beta_P x + \mu_P \, x \, y} \, \xi_x(t), \nonumber \\
    \dot{y} & = \beta_T \, x - \mu_T \, y + \sqrt{\frac{1}{V}} \sqrt{ \beta_T \, x + \mu_T \, y} \, \xi_y(t),
\end{align}
where $\xi_y(t)$, $\xi_x(t)$ are Gaussian white noises with $\langle\xi_y(t) \xi_x(s)\rangle=0$, and $\langle\xi_x(t) \xi_x(s)\rangle=\langle\xi_y(t) \xi_y(s)\rangle=\delta(t-s)$. 

The equation for the average densities is obtained from Eq.~\eqref{eq:general_SDE} by neglecting the noise terms and will be discussed in more detail in Sec.~\ref{sec:deter}. We advance here that the resulting deterministic system admits two stationary solutions: the absorbing state, $x_0 = y_0 = 0$, and the coexistence state, $x^* = \beta_p \mu_T / (\beta_T \mu_P)$ and $y^* = \beta_P / \mu_P$.
In contrast, in the full stochastic description given by Eq.~\eqref{eq:general_SDE}, the only true stationary state is the absorbing point $(x_0, y_0)$. Once the system reaches this state, fluctuations cease entirely, since no stochastic events can drive it away. Moreover, while in the deterministic limit the approach to the absorbing state would occur only asymptotically (i.e., at infinite time), in the stochastic dynamics absorption is reached in finite time due to fluctuations~\cite{Hinrichsen2000}.

Without loss of generality, we consider the following change of variables, which simplifies the parametric analysis of the equations: $x' = \beta_T\mu_P/(\beta_P\mu_T)\,x$, $y' = \mu_P/\beta_P \, y$, $t' = \beta_P\, t$ and consider the noise scaling $\xi(t)=\sqrt{\beta_P}\,\xi(t')$. Dropping down the primes, we obtain the following adimensional Langevin equation:
\begin{align}\label{eq:full_Langevin}
    \frac{\partial x}{\partial t} & = x\,(1-y)+ D_x \sqrt{x\,(1-y)} \, \xi_x(t), \nonumber \\
    \frac{\partial y}{\partial t} & = \rho( x -  y) + D_y \sqrt{x + y} \, \xi_y(t),
\end{align}
where $\rho = \mu_T/\beta_P$ describes the ratio of the timescales of $x$ and $y$, and $D_x = \sqrt{\beta_T\,\mu_P/(\beta_P\,\mu_T\,V)}$ and $D_y = \sqrt{\mu_P\,\mu_T/(\beta_P^2\,V)}$ describe the the noise strength on the $x$ and $y$ variables.
\subsection{Neglecting toxin concentration fluctuations}

As discussed in Sec.~\ref{sec:Microscopic_derivation}, demographic fluctuations arise from the discrete nature of individual agents~\cite{VanKam}, which reproduce and die at random times and therefore generate stochastic dynamics. However, stochastic effects may become negligible in appropriate limits, for instance, in highly diluted systems (see Sec.~\ref{sec:deter}). To quantify the relative importance of concentration fluctuations in the population and in the toxin field, we introduce the diffusion ratio
\begin{equation}
    \alpha = \frac{D_y}{D_x} = \sqrt{\frac{\rho}{r}},
    \label{Eq:alpha_rate}
\end{equation}
where $r = \beta_T / \mu_T$ is the ratio between the toxin production rate and the toxin degradation rate. This quantity measures the typical amount of toxin accumulated over its lifetime.
We argue that autotoxic systems displaying boom--bust dynamics are naturally associated with small diffusion ratios ($\alpha \ll 1$). First, as discussed in Sec.~\ref{sec:deter}, oscillatory behavior in the deterministic dynamics requires $\rho \leq 4$, placing the system in a regime of slow environmental feedback and implying $\alpha \le 2/\sqrt{r}$. Second, empirical studies indicate that allelochemicals are often produced continuously and in large quantities, while degrading on timescales ranging from weeks to months or longer, depending on soil conditions and microbial activity~\cite{Inderjit2003, Blum2011, hierro_ecological_2021}. This combination of high production rates and comparatively slow degradation leads to large values of $r$. Indeed, using data from yeast fermentation experiments exhibiting autotoxic effects~\cite{Pagliardini2010,deBruyn2020,Labedz2017,Sahana2024}, we estimate $r \sim 10^{13}$ for this example (see App.~\ref{Apx:Parameters}).

Small diffusion ratios imply that toxin concentrations fluctuate weakly compared to population fluctuations. Intuitively, when toxin production events are very frequent (large $r$), their stochastic contributions average out, leading to effectively deterministic toxin dynamics. This observation allows us to simplify the analysis by neglecting fluctuations in the toxin variable $y$. Under this approximation, the dynamics reduce to
\begin{align}
    \frac{\partial x}{\partial t} & = x - x y + D \sqrt{x + x y}\, \xi(t), \nonumber \\
    \frac{\partial y}{\partial t} & = \rho (x - y),
    \label{Eq:langevin}
\end{align}
with only two parameters: $\rho = \mu_T / \beta_P$, describing the timescale separation between $x$ and $y$, and $D = D_x = \sqrt{\beta_T \mu_P / (\beta_P \mu_T V)}$, which controls the intensity of demographic fluctuations.

\subsection{Mean field Deterministic dynamics}
\label{sec:deter}

In this section, we analyze the deterministic mean-field model obtained by neglecting the noise terms in Eq.~\eqref{eq:full_Langevin} (i.e., setting $D_x = D_y = 0$). This approximation corresponds to the limit of a highly diluted system ($V \to \infty$). This description provides a baseline of the underlying dynamics, against which the effects of finite fluctuations can later be assessed. The resulting system of ordinary differential equations (ODEs) reads
\begin{align}
    \frac{dx}{dt} &= x - x\,y, \nonumber
    \\
    \frac{dy}{dt} &= \rho(x - y).
    \label{Eq:ODEs_simpl}
\end{align}

This deterministic system exhibits two fixed points: an unstable, unpopulated state $S_0 = (0,0)$ and a stable, populated state $S_p = (1,1)$, which is the sole attractor of the system.

The transition from low-population initial conditions to $S_p$ is controlled with $\rho$. For large values of $\rho$, this transition is monotonic. 
In fact, in the limit $\rho \rightarrow \infty$, the fast dynamics of the inhibitor can be adiabatically simplified, recovering the logistic equation $dx/dt = x(1 - x)$. In this limit, the toxin dynamics are effectively reduced to a direct interaction between individuals in the population, as the transient concentration of toxins becomes negligible.

\begin{figure*}
    \centering
    \includegraphics[width=1\textwidth]{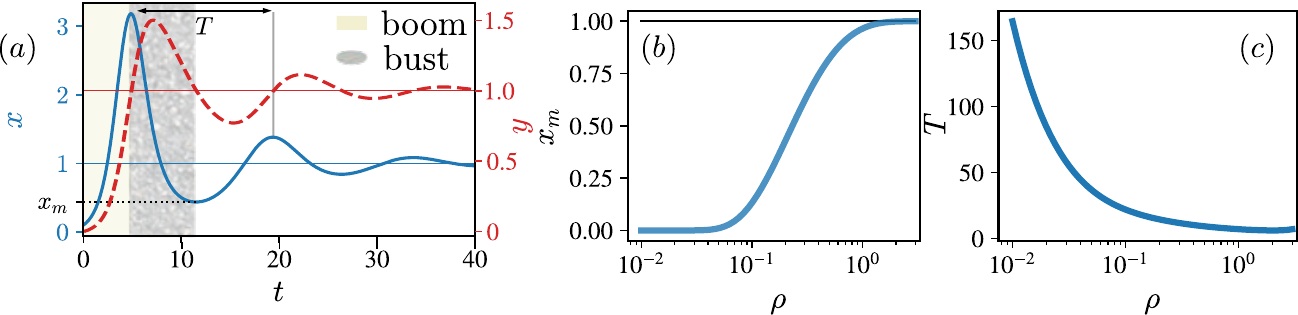}
    \caption{\textbf{Deterministic behavior.} (a) Example of a deterministic trajectory exhibiting a boom--bust cycle (highlighted by shaded rectangles) followed by oscillations around the coexistence fixed point (horizontal solid line) for $\rho = 0.2$. The $x$ and $y$ axes share a common zero-baseline. The values of both variables at the stable equilibria $S_P$ are shown with horizontal lines. The figure shows the quantities plotted in (b) and (c).
(b) Minimum population value after the initial boom--bust cycle, which tends to $S_0$ as $\rho$ decreases. For large $\rho$, oscillatory behavior fades out, and the minimum coincides with the stable populated fixed point $S_p$, indicated by a horizontal solid line.
(c) Time interval between the first two peaks in population density.}
    \label{Fig:deterministic}
\end{figure*}

As $\rho$ decreases, the approach to $S_p$ becomes oscillatory for $\rho < 4$.
The onset of oscillations occurs at $\rho = 4$, where the two real negative eigenvalues of the fixed point become a pair of complex conjugate eigenvalues with negative real part, and therefore the fixed point $S_p$ transitions from a stable node to a stable focus. This is known in the literature as a Belyakov--Devaney pseudobifurcation~\footnote{This transition is a pseudobifurcation instead of a bifurcation because the stability of the solution does not change in the process.}\cite{Homburg2010}.
Fig.~\ref{Fig:deterministic}(a) illustrates the oscillatory deterministic trajectories. As $\rho$ decreases, both the amplitude and the period of these oscillations grow, and the trajectories approach the unpopulated equilibrium $S_0$. This trend is shown in Figs.~\ref{Fig:deterministic}b–c: near $\rho=0$, the dynamics slow markedly in the vicinity of $S_0$, causing the interval between successive oscillations ($T$) to diverge and the minimal concentration after the first boom ($x_m$) to approach zero. Notice in Fig.~\ref{Fig:deterministic}(a) that toxin values remain relatively large during most of this close to extinction transient.

\section{Results}
\label{Sec:2}

\subsection{Extinction pathways}

In this section, we return to the stochastic trajectories of Eq.~\eqref{Eq:langevin}. Since, as far as the authors know, Eq.~\eqref{Eq:langevin} cannot be solved analytically, our methods rely on sampling trajectories numerically with Milstein's algorithm, which is a standard approach for the efficient integration of stochastic differential equations with demographic noise~\cite{Toral2014}. In particular,  we study trajectories with toxin-free initial conditions ($y(t=0) = 0$), small initial populations ($x(t=0)<1$), and in the regime with $\rho<4$. As already presented in Section~\ref{sec:deter}, for these parameter values, deterministic trajectories, i.e., with $D=0$, would converge after some oscillations to the stationary steady state $S_P = (1,1)$. However, in a stochastic scenario ($D>0$), the long-term limit $t\to\infty$ is identical for all realizations: the system eventually falls into the absorbing manifold $x=0$ due to random fluctuations. Once the population is extirpated, the toxin concentration $y$ decays deterministically toward the final state $S_0 = (0,0)$.

\begin{figure*}
    \centering
    \includegraphics{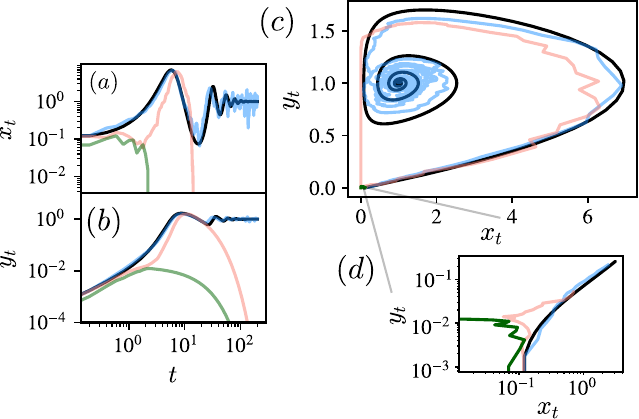}
    \caption{\textbf{Stochastic behavior: extinction pathways.} Instances of trajectories for fixed $\rho = 0.08$ and $D=0$ (black), $D=0.1$ (blue), $D=0.2$ (orange), and $D=0.4$ (green). Panels (a) and (b) show the components $x$ and $y$, respectively, of the trajectories as functions of time in log-log scale. Panel (c) shows trajectories in the phase space, while zoom in close to the (0,0) is shown in (d) in log-log scale. Different extinction pathways are observed as $D$ changes. For $D=0$ the trajectory follows the deterministic path and no extinction occurs (black line). For small noise ($D=0.1$, blue curve), stochastic trajectories fluctuate about the deterministic path ($D=0$), reaching a meta-stable state and experiencing extinction through a rare fluctuation at very long times. For intermediate noise strength ($D=0.2$, orange curve), trajectories are likely to touch the absorbing state after reaching the macroscopic maximum. Extinction times in this phase will be of the order of the time to reach the first minimum of the deterministic trajectory. For big noise ($D=0.4$, green curve), trajectories are likely to become extinct before reaching the first maximum of the deterministic trajectory. Trajectories were obtained integrating Eq.~\eqref{Eq:langevin} using Milstein algorithm with discretization $\Delta t=10^{-4}$ and initial conditions $x_0=0.1$, and $y_0=0$. }
    \label{Fig:trajectories_different_Ds}
\end{figure*}

Although eventual extinction is certain, the path to the absorbing state varies dramatically between realizations. Trajectories range from almost immediate collapse (green paths in Fig.~\ref{Fig:trajectories_different_Ds}) to ``persistent'' survival, where the trajectory fluctuates around $S_P$ for durations exceeding observational timescales~\cite{Meerson2009} (blue paths in Fig.~\ref{Fig:trajectories_different_Ds}). While the existence of different dynamical regimes is evident qualitatively simply by looking at the trajectories (see illustration in Fig.~\ref{Fig:trajectories_different_Ds}), defining what constitutes ``immediate collapse'' or ``persistent'' behavior rigorously enough to allow quantitative analysis is tricky in stochastic models.

To tackle this problem, we first define the winding phase $\theta_W$, which measures the total accumulated counterclockwise angle of the trajectory around $S_P$ from the initial state to absorption at $t\rightarrow \infty$:
\begin{equation}\label{eq:def_winding_angle}
    \theta_W = \int_0^\infty d\theta(t)
    = 2\pi N_W-\Delta\theta,
\end{equation}
where $\theta(t)$ is the unwrapped (continuous) angle of the trajectory around $S_P$\footnote{Notice that the unwrapped angle can be described as function of the system variables as $d\theta=\frac{(x-1)dy-(y-1)dx}{(x-1)^2+(y-1)^2}$.}, the winding index $N_W$ is an integer random variable which describes the number of complete contra-clockwise revolutions around $S_P$ over the interval $t\in[0,\infty)$, and $\Delta\theta=\theta(0)+\frac{3\pi}{4}$ is the difference between the initial angle $\theta(0) = arg(x(0)-1+i(y(0)-1))$ and the final angle when the trajectory reach $S_0$ for infinite times. Notice that, as the dynamics in $y$ are deterministic, negative winding indices are impossible in our setup \footnote{For $x>y$ (resp.\ $x<y$) one necessarily has $dy/dt>0$ (resp.\ $dy/dt<0$), implying that the dynamics around $S_p$ must be counterclockwise.}.

Thus, the winding index, $N_W = (\theta_W+\Delta\theta)/2\pi$, allows the classification of  trajectories in three groups (see Fig.~\ref{Fig:trajectories_different_Ds}):
\begin{itemize}
    \item \textbf{Short-lived trajectories} ($N_W=0$). Trajectories that are absorbed before completing a single boom-bust excursion.
    \item \textbf{Excitable trajectories} ($N_W = 1$). Trajectories that undergo a full boom-bust cycle but are absorbed during the subsequent ``bust'' phase due to high toxin levels.
    \item \textbf{Persistent trajectories} ($N_W > 1$). Persistent trajectories are those that survive beyond the first boom–bust cycle. This regime spans a broad range of characteristic absorption times—from trajectories absorbed after only a few peaks to ones that fluctuate around $S_P$ for exponentially long durations where extinction is due to rare fluctuations~\cite{Aguilar2022_rare}.
\end{itemize}

\subsection{Noise induced excitabilty}

A key feature of deterministic excitable systems is the presence of a threshold in the initial condition: trajectories starting below this threshold decay to the rest state without excitation, whereas those starting above undergo a large excursion in phase space before returning back to the rest state. In our stochastic model, this rigid boundary is replaced by a probabilistic landscape. Due to stochastic fluctuations, identical initial conditions and parameters can result in different pathways to the absorbing state, corresponding to different values of the winding index $N_W$. We therefore, have to characterize the system through the probability of obtain each pathway to extinction, with $p_{sl} = \text{Prob.}(N_W=0)$, $p_e = \text{Prob.}(N_W=1)$, and $p_p = \text{Prob.}(N_W>1)$ representing the probabilities of having shot-lived, excitable and persistent trajectories respectively. These probabilities are conditioned on parameter values and initial conditions.

\begin{figure*}
    \centering
    \includegraphics[width=1\linewidth]{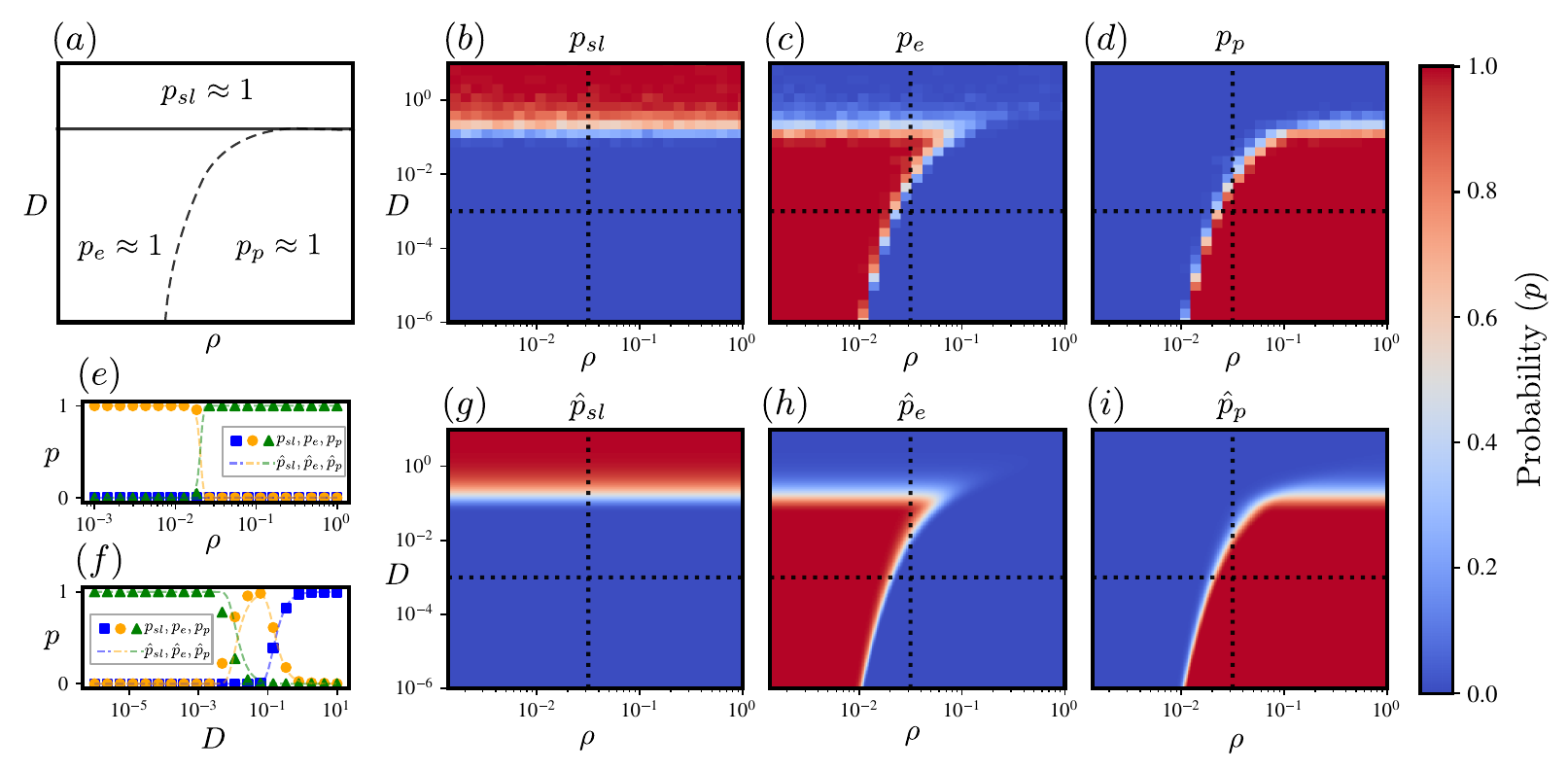}
    \caption{\textbf{Extinction probabilities} In (a), sketch regions in parameter space where absorption probability is dominated by short-lived trajectories ($p_{sl}$), excited trajectories ($p_e$), and persistent trajectories ($p_p$). Panels (b–d) show, respectively, the probabilities of extinction via short-lived, excited, and persistent trajectories as functions of $\rho$ and $D$.  Panels (e) show the numerically computed extinction probabilities and analytical approximation for fixed $D=10^{-3}$ as a function of $\rho$. Panel (f) shows probabilities and approximations for fixed $\rho=10^{-1.5}$ as a function of $D$. Panels (g-i) show the same information as (b-d), but using our approximated analytical probabilities [Eq.~\eqref{Eq:Anal_prob}], instead of Monte Carlo estimators. In all cases we set $x(0)=0.01$ and $y(0)=0$.}
    \label{fig:x_th_D_rho}
\end{figure*}

In Fig.~\ref{fig:x_th_D_rho},  we show the probabilities of observing each extinction scenario as functions of $D$ and $\rho$, with $y_0=0$ and $x_0 = 0.01$ held fixed. Three distinct regimes emerge with remarkably sharp transitions, each dominated by a different extinction pathway (see sketch in Fig.~\ref{fig:x_th_D_rho}(a). Short-lived trajectories dominate at high noise intensity $D$, where demographic fluctuations drives the system to extinction before the first ``boom'' can occur (Fig.~\ref{fig:x_th_D_rho}(b). Conversely, when $D$ is small, the outcome is also governed by the timescale separation $\rho$. For small $\rho$ (slow toxin decay), the system is dominated by excitable trajectories (Fig.~\ref{fig:x_th_D_rho}(c)), whereas for large $\rho$ is dominated by persistent trajectories (Fig.~\ref{fig:x_th_D_rho}(d)). This transition occurs because, as $\rho$ decreases, the deterministic drift makes the trajectories to visit states closer to extinction ($x=0$) and to stay there for longer periods of time. Small fluctuations can then very easily kill the process.

In order to better understand the underlying physics of the results shown in Fig.~\ref{fig:x_th_D_rho}, we derived analytical approximations of the absorption probabilities (see Appendices~\ref{apx:analytic1} and \ref{apx:analytic2}):
\begin{align}
    \hat{p}_{sl}(\rho,D,x_0) &= e^{-x_02D^{-2}}, \nonumber \\
    \hat{p}_e(\rho,D,x_0) &= (1-\hat{p}_{sl})\,e^{-\eta(\rho)\, 2D^{-2}} ,\nonumber \\
    \hat{p}_{p}(\rho,D,x_0) &= 1 -\hat{p}_{sl} -\hat{p}_e,
    \label{Eq:Anal_prob}
\end{align}
where the term $\eta(\rho)$ represents somehow the ``barrier height'' that separates the trajectory from the absorbing state during the first bust. The value of $\eta(\rho)$ is given by the following expression:
\begin{align}
    \eta(\rho)&=\frac{-e^{-y_c/\rho}y_c^{1/\rho}}{ \rho^{1/\rho}\left[\frac{1}{\rho}\,\gamma\left(\frac{1}{\rho},z\right)+\gamma\left(\frac{1}{\rho}+1,z\right)\right]^{z=y_c/\rho}_{z=1/\rho}}
\label{Eq:eta}
\end{align}
with $\gamma(s,a)$ being the incomplete lower gamma function and $y_c=y_c(\rho)$ being the value of $y$ at which the unstable manifold of $S_0$ cuts for first time $x=1$ with $y>1$. For small values of $x_0$, $y_c$ is a good approximation of the cut of the deterministic trajectories with $x=1$. Notice that $y_c$ depends only on the value of $\rho$, as it is the only parameter of the deterministic dynamics given by Eq.~\eqref{Eq:ODEs_simpl}. The agreement between the probabilities predicted by Eq.~\eqref{Eq:Anal_prob} and results from simulations is shown in Fig.~\ref{fig:x_th_D_rho} (e-i).

Remarkably, extinction pathways are governed by exponential distributions balancing a variety of characteristic scales. The emergence of such exponential forms explains the abrupt transitions observed in Fig.~\ref{fig:x_th_D_rho}. Despite the complexity of Eq.~\eqref{Eq:Anal_prob}, the role of the diffusion constant $D$ is quite transparent. The scale $\ell_d = D^2/2$ sets the typical fluctuation scale. The competition between this noise magnitude $\ell_d$ and the population density scales $x_0$ and $\eta(\rho)$ will determine the excitation threshold of the system and the transition between persistence and excitation regimes, respectively.

The excitation threshold is controlled by the ratio $x_0/\ell_d$. When $x_0 \ll \ell_d$, fluctuations dominate at short times, and most trajectories are short-lived ($p_{sl}\sim1$). When $x_0\gg\ell_d$, the population will almost certainly grow for small times and a ``boom'' will be observed ($p_e+p_p\sim 1$). Therefore, the transition between absorbed and non-absorbed trajectories shows threshold-like features typical of excitable systems, and the location of this threshold depends, at least in the analytical approximation, exclusively on the diffusion parameter. This threshold might be understood in a probabilistic way; failed excitations above the threshold, and unexpected booms below it are both possible, although unlikely, outputs of the model. Notice that the excitation threshold is independent of $\rho$, at least for the small values of $x_0$ considered in this work.

Non-short-lived trajectories will either be excitable or persistent. The probability of obtaining one or the other depends on the ratio $\eta(\rho)/\ell_d$. During the first bust, if $\eta(\rho) \ll \ell_d$, the trajectory is almost certainly absorbed and results in an excitable trajectory ($p_e \gg p_{sl}$). Instead, when $\eta(\rho)\gg \ell_d$, the trajectory will almost certainly survive the first bust and become a persistent trajectory ($p_e\ll p_p$). Notice that, at least in the analytical approximation, the excitation-persistence transition is independent of the initial conditions, and driven only by the parameters $D$ and $\rho$.

The distinction between excitable and persistent trajectories is again governed by the balance between two characteristic scales: the fluctuation scale $\ell_d(D)$ and the typical distance to absorption after the ``bust'', $\eta(\rho)$. The scale $\eta$ depends only on the ratio between population and toxins timescale, $\rho$, and thus encapsulates the deterministic system dynamics. In particular, it can be shown that for any non-zero $D$ small enough values of $\rho$ make the system fall in the excitable regime (see App.~\ref{Appx:AsymExp}).

\section{Discussion}
\label{Sec:3}
The stochastic approach allows us to deal with the unpredictability of populations at the individual level. In this framework, this unpredictability translates into population fluctuations that can be related to the microscopic processes of the system. Rather than reducing the explanatory power of ecological models, this randomness accounts for macroscopic features observed in real ecosystems, such as sudden extinction events. We have demonstrated that noise-induced excitability arises naturally in a simple autotoxicity model when demographic fluctuations are incorporated. To achieve similar ``boom-bust-extirpation'' behavior in deterministic models, one typically must introduce explicit positive feedback mechanisms, such as a strong Allee effect \cite{MorenoSpiegelberg2025}. Our results suggest that such behaviors can instead emerge from the interplay between negative feedback and noise, reducing the number of assumptions required to model excitable ecological dynamics.

In this paper, we utilize the standard Van Kampen's system expansion technique to derive a mesoscopic Langevin description from individual-level first principles. Our subsequent approximation, neglecting toxin fluctuation while keeping population noise, is grounded on the analysis of the typical scales of the problem and can be applied to arbitrary autotoxic systems.

The resulting system, in its adimensional form, is governed by two parameters: the time scale ratio between toxin and population dynamics, $\rho$; and the noise intensity $D$. The parameter $\rho$ dictates the lag in environmental feedback; small $\rho$ indicates slow toxin dynamics, leading to large population outbreaks followed by deep, prolonged ``busts''. The noise intensity $D$ scales with the inverse of the square root of the system size, i.e., the carrying capacity of the system. In large systems (small $D$), fluctuations are negligible, and trajectories follow the deterministic dynamics very closely. For small systems (big $D$), fluctuations dominate, and the time of absorption (extinction) decreases significantly. Nevertheless, due to the stochastic nature of the model, different outcomes can be obtained from the same set of parameters and initial conditions.

We classified trajectories into three distinct extinction pathways: short-lived, excitable, and persistent. The probability of a system following a specific kind of trajectory is a function of the initial density and the model parameters. We have analyzed these probabilities for small initial populations without initial toxin concentrations, using numerical simulations and analytical approximations.

We have identified a probabilistic threshold for the initial conditions. Populations starting below this threshold are likely to be extirpated before showing a first boost (short-lived), while those above are very likely to show excitable or persistent behavior. Furthermore, we have observed a sharp transition in the parameter space between a region where most of the non-short-lived trajectories are excitables, and a region where most of the non-short-lived trajectories are persistent. This transition to an excitable regime differs from those known for deterministic excitability, as it does not involve the destruction of a large-amplitude limit cycle \cite{Izhikevich2006}, suggesting that the behavior studied in this article represents a previously unexplored form of excitability.

\section{Conclusions and perspectives}
\label{Sec:5}
From a biological point of view, our results suggest that species with slow toxin dynamics or those introduced to small habitats are significantly more prone to extinction after a boom-bust event. Furthermore, as the excitable-persistent transition is really sharp, a small change in the environmental condition or system size can dramatically change the statistical fate of the population. When time-series data from real ecosystems are available, our framework could be used to assess extinction probabilities through inference techniques that allow parameter estimation from time-series data~\cite{aguilar2025limits}. Testing the different phases of extinction probabilities empirically remains an important direction for future work; this could be feasible in microscopic ecology experiments by setting different initial conditions for microbial populations in controlled environments to compute empirical extinction probabilities.

In conclusion, demographic fluctuations do not merely blur deterministic dynamics; they generate a well-defined probabilistic structure that organizes extinction pathways. By identifying the characteristic fluctuations and dynamical scales that control absorption probabilities, we provide a framework where excitability emerges as a noise-driven dynamical phase. Beyond autotoxic systems, our approach offers a general route to studying boom-bust-extirpation dynamics by other population-control mechanisms such as resource depletion \cite{aguilar2025unraveling}, pathogen proliferation \cite{Stricker2016}, or predation feedbacks \cite{Keane2002,Carlsson2011}. Furthermore, the mechanism of noise-induced excitability may find applications in broader contexts, including viral clearance after acute infection, the eradication of epidemics \cite{vanHerwaarden1997}, the study of spatiotemporal patterns in vegetation \cite{Vincenot2017} or the elimination of cancerous cells in cancer treatment \cite{Lopez2019,Ramirezvaila2023}.
\section*{Data Availability}
The data that support the findings of this study are available from the corresponding author upon reasonable request.

\section*{Code availability}

The code used to generate the plots of this study is available in ref.~\cite{github_link}.

\section*{Acknoledgments}
P.M.-S. acknowledges financial support from project CYCLE (PID2021-123723OB-C22) funded by MCIN/AEI/10.13039/501100011033 and the European Regional Development Fund (ERDF) ``A way of making Europe'', the María de Maeztu project CEX2021-001164-M funded by the MCIN/AEI/10.13039/501100011033, and the European Union’s Horizon’s 2020 research and innovation programme (Grant agreement ID: 101093910, Ocean Citizen). J.A. acknowledges financial support under the National Recovery and Resilience Plan (NRRP), Mission 4, Component 2, CUP
2022WPHMXK, Investment 1.1, funded by the European Union – NextGenerationEU – Project Title:  ``Emergent Dynamical Patterns of Disordered
Systems with Applications to Natural Communities''.
The authors acknowledge the use of generative AI for language editing and to improve the clarity and readability of the manuscript. After using these tools, the author reviewed and edited the content as needed and takes full responsibility for the content of the published article.
\appendix
\section{Parameter estimation}
\label{Apx:Parameters}
In this appendix, we introduce the parameter estimation for yeast fermentation and seagrass meadows.

In anaerobic conditions, yeast generates alcohol as a waste product in its metabolism. This alcohol, once accumulated in the medium, inhibits yeast growth and even causes the culture to collapse. The parameters for these kind of system are very dependent on the environmental condition, but to take our selfs an idea of the magnitude of the parameters lets take a growth rate of $\beta_P=0.35\, h^{-1}$ \cite{Pagliardini2010}, and a decay rate of ethanol of $\mu_T=7.8 \; 10^{-2} \, h^{-1}$ \cite{deBruyn2020}. The production of ethanol can be considered as $1.32\, g \, g^{-1}_{DCW}h^{-1}$ \cite{Pagliardini2010}, considering a mass of a single yeast cell of $47.67\,pg_{DCW}$ \cite{Labedz2017} and the molar mass of ethanol ($46.07 g\;mol^{-1}$) we obtain a ethanol production rate $\beta_T = 8.16\;10^{11}h^{-1}$. The mortality rate of the yeast due to ethanol can be estimated from the ethanol concentration where growth of the population stops, around $12^{\circ}$ \cite{Sahana2024}, i.e. $2.06\,mol\,L^{-1}$. As for this concentration reproduction and mortality is the same, we can obtain $\mu_P = 2.82 \, 10^{-25}\,L\,h^{-1}$.

Computing the magnitude of the noise in this case, we obtain $D_x=2.90 \,10^{-6}\;V^{-1/2}$ and $D_y = 4.23 \;10^{-13}\;V^{-1/2}$. Notice that $D_y \ll D_x$. Similar results are expected in systems where the toxic concentration of toxins is of the order of $1\; mol$ per volume and the toxin production is of the order of the Avogadro number. Notice that in these cases, the concentration of toxic molecules is many orders of magnitude above the number of individuals, and the fluctuations in the toxins can be neglected.

\section{Computation of absorption probability in the limit $\rho\to0$.}
\label{apx:analytic1}
In the limit $\rho\to 0$ and assuming that toxins are absent at time zero ($y(t=0)=0$), Eq.~\eqref{Eq:langevin} simplifies to a one–dimensional equation:
\begin{equation}\label{eq:Branching_process}
    \dot{x}= x + D\sqrt{x}\,\xi(t).
\end{equation}
This equation is known as a branching process whose statistics are all analytically known, since the associated Fokker–Planck equation,
\begin{align}
    \partial_t p(x\mid D, x_0,\gamma) = -\partial_x \big( x\, p(x\mid D, x_0,\gamma)\big) + \frac{D^2}{2}\,\partial_{xx}\big( x\,p(x\mid D, x_0,\gamma)\big),
\end{align}
can be solved exactly~\cite{Feller1951}. In particular, the probability $\pi_0$ that the process in Eq.~\eqref{eq:Branching_process} with initial condition $x(t=0)=x_0\ge0$ is absorbed at $x=0$ before reaching $x=a>x_0$ satisfies~\cite{VanKam}
\begin{align}
     x_0 \partial_{x_0}\pi_0(x_0)+\frac{D^2}{2}\, x_0
\partial^2_{x_0}\pi_0(x_0)=0,
\end{align}
with boundary conditions
\begin{align}
    \pi_0(0)&=1, \nonumber\\
    \pi_0(a)&=0.    
\end{align}
The solution consistent with these boundary conditions is
\begin{align}\label{eq:absorption_probability_small_rho}
    \pi_0(x_0,a)=1-\frac{e^{-2 x_0/D^2}-1}{e^{-2a/D^2}-1}.
\end{align}
Because Eq.~\eqref{eq:Branching_process} lacks a saturating nonlinear term, the process is unbounded. Thus, when $x(t=0)=x_0\approx 0$, trajectories either get absorbed at $x=0$ at early times or diverge, with $\lim_{t\to\infty} x(t)=\infty$. In this regime, the probability of absorption at $x=0$ can be approximated by the probability of observing a short–lived trajectory:
\begin{equation}
    \lim_{\rho\to 0} p_{sl} \approx \lim_{a\to\infty} \pi_0(x_0,a)= e^{-2x_0/D^2}.
\end{equation}
Because population and toxin growth are unsynchronized at small $\rho$, trajectories that are not short–lived tend to diverge (see Fig.~\ref{fig:x_th_D_rho}), and the probability of persistent trajectories vanishes. Therefore,
\begin{equation}
    \lim_{\rho\to0} p_e = 1-e^{-2x_0/D^2}.
\end{equation}

\section{Computing probability of excitability in the limit $D\to 0$}
\label{apx:analytic2}

Following the paper on epidemics\cite{vanHerwaarden1997}, we can write the equation for the probability of getting absorbed in $x=0$ ($u(x,y)$) as:
\begin{equation}
    x(1-y)\partial_xu+\rho(x-y)\partial_yu+\frac{1}{2}D^2x(1+y)\partial_{xx}u=0,
\end{equation}
with boundary conditions $u(0,y)=1$ for any $y$ and $u(1,y)=0$ for $y<1$. Considering the change of variables $\zeta=D^{-2}x$ and considering the limit of small noise $D \rightarrow 0$ we obtain the equation:
\begin{equation}\label{eq:eq_prob_u}
    \zeta \, (1-y) \,\partial_\zeta \,u-\rho \,y\,\partial_y u +\frac{1}{2}\, \zeta\, (1+y)\, \partial_{\zeta\zeta}\,u=0.
\end{equation}
Notice that in this equation the drift component in $y$ does not depend on $\zeta$, considering a decoupled evolution in the toxin density for low populations.

We propose the change of variables
\begin{equation}
    \xi = \xi\left(\zeta,y\right),\quad\quad \nu = \nu\left(\zeta,y\right).
\end{equation}
Introducing the change of variables in Eq.~\eqref{eq:eq_prob_u}, grouping the terms by derivatives of $\xi$ and $\nu$, and imposing the form for the equation after the change of variables,
\begin{equation}
    \partial_\nu \,u = \xi\,\partial^2_\xi u,
    \label{eq:absortion_simpl}
\end{equation}
we find the following equations for $\xi$ and $\nu$,
\begin{equation}
    \zeta \,( 1-y)\,\partial_\zeta\,\xi-\rho \,y \,\partial_y\,\xi+\frac{1}{2}\zeta (1+y)\partial^2_\zeta \xi=0,
\end{equation}
Looking for solutions with $\partial^2_\zeta \xi=0$ we find
\begin{equation}
    \xi=\zeta\,e^{-\frac{y}{\rho}}y^{\frac{1}{\rho}}.
\end{equation}
The equation from the coefficient of $\partial_\xi u$ reads,
\begin{equation}
    \frac{1}{2}\zeta \,( 1+y)\,\left(\partial_\zeta\,\xi\right)^2 = \frac{1}{2} \xi(1+y)e^{-\frac{y}{\rho}}y^{\frac{1}{\rho}}.
\end{equation}
Imposing solution without the term with $\partial^2_\nu$ we impose
\begin{equation}
\frac{1}{2}\zeta(1+y)\nu_\zeta^2=0
\end{equation}
obtaining $\nu(\zeta,y)=\nu(y)$. And finally to obtain the term $\xi\partial_{\xi \xi}u$ we impose


\begin{equation}
\nu_y = -\frac{1}{2\rho}(1+y)e^{-y/\rho}y^{1/\rho-1}.
\end{equation}.

This way, we obtain the following new variables :
\begin{align}
    \xi(\zeta,y)&=\zeta e^{-y/\rho}y^{1/\rho} \\
    \nu(\zeta,y)&=-\frac{1}{2\rho}\int^y_1 (1+s)e^{-s/\rho}s^{1/\rho-1}ds = \nonumber\\ 
    &=- \frac{\rho^{1/\rho}}{2}\left[\frac{1}{\rho}\,\gamma\left(\frac{1}{\rho},z\right)+\gamma\left(\frac{1}{\rho}+1,z\right)\right]^{z=y/\rho}_{z=1/\rho},
\end{align}
where we have used the lower incomplete gamma function,
\begin{equation}
    \gamma(a,x) =\int^x_0 s^{a-1}e^{-s} \,ds.
\end{equation}

The Eq.~\eqref{eq:absortion_simpl} get as solution the function:
\begin{equation}
    u(\xi,\nu)=e^{-\xi/\nu}
\end{equation}
for boundary condition $u(0,\nu)=1$ and $u(\xi,0)=0$, which corresponds in $x$ and $y$ with $u(0,y)=1$ and $u(x,1) = 0$. We define the expression $\eta = \xi/\nu$ as shown in Eq.~\eqref{Eq:eta}.

\section{Asymptotic expansion of $\nu(y)$ for $\rho\to0^+$ and $D\ne0$}
\label{Appx:AsymExp}

We consider the denominator of Eq.~\eqref{Eq:eta}:
\begin{equation}\label{eq:nu-def}
\nu(y)=-\frac{\rho^{1/\rho}}{2}\;\Big[\tfrac{1}{\rho}\,\gamma\!\big(\tfrac{1}{\rho},z\big)+\gamma\!\big(\tfrac{1}{\rho}+1,z\big)\Big]_{z=1/\rho}^{z=y/\rho},
\end{equation}
where $\gamma(\cdot,\cdot)$ is the lower incomplete gamma function and $0<\rho\ll1$.

Introduce the notation
\begin{equation}
a:=\frac{1}{\rho}>0\;\;\text{(so $a\to\infty$ as $\rho\to0^+$)},
\end{equation}
and define
\begin{equation}
\mathcal S(z):=\tfrac{1}{\rho}\,\gamma\!\big(\tfrac{1}{\rho},z\big)+\gamma\!\big(\tfrac{1}{\rho}+1,z\big)
= a\,\gamma(a,z)+\gamma(a+1,z).
\end{equation}
Using the standard recurrence
\begin{equation}
\gamma(a+1,z)=a\,\gamma(a,z)-z^{\,a}e^{-z},
\end{equation}
we obtain the exact simplification
\begin{equation}\label{eq:S-simplify}
\mathcal S(z)=2\,a\,\gamma(a,z)-z^{\,a}e^{-z}.
\end{equation}
Substituting \eqref{eq:S-simplify} into \eqref{eq:nu-def} and simplifying yields
\begin{equation}\label{eq:nu-exact}
\nu(y)=-a^{1-a}\big[\gamma(a,a\,y)-\gamma(a,a)\big]
\;+\;\tfrac12\big(y^a\,e^{-a\,y}-e^{-a}\big). 
\end{equation}
Where in the limit $a\rightarrow \infty$ the second right side terms decay exponentially to zero, and the lower incomplete gamma functions are asymptotically equivalent to:
\begin{align}
    \gamma(a,a) \sim \frac{1}{2}\Gamma(a) \qquad\gamma(a,ya)\sim \Gamma(a) \qquad (y>1),
\end{align}
when $y>1$ \cite{dlmf8.12}. And therefore:
\begin{equation}
    \nu(y)\sim -\frac{1}{2}a^{1-a}\Gamma(a)\sim\sqrt{\frac{\pi \,a}{2}}e^{-a}
\end{equation}

Notice that using Stirling's approximation for the gamma function for big values of $a$, i.e. $\Gamma(a)\sim\sqrt{2\pi}\, a^{a-1/2}\, e^{-a}$, and substituting in (\ref{eq:nu-exact}) we found:
\begin{equation}
    \nu(y)\sim -\sqrt{\frac{\pi}{2\,\rho}}\;e^{-1/\rho},\qquad(0<\rho\ll1,\;y>1)\,.
\end{equation}

Using this expresion in Eq.~\eqref{Eq:Anal_prob} we obtain that for $\rho \ll 1$:
\begin{equation}
    \ln(\hat{p}_e/(1-\hat{p}_{sl}))\sim \sqrt{\frac{2}{\pi}}\frac{x}{D^2}\rho^{1/2}e^{(ln(y)-y+1)/\rho},
\end{equation}
which converges to zero in the limit $\rho \rightarrow 0$ if $D\neq0$. 

The physical interpretation of this result is that for any non-zero value of $D$, most of the not short-lived trajectories would be excitable if $\rho$ is low enough.

\bibliography{sample_abbreviate}

\end{document}